\documentclass[epj]{svjour}

\usepackage{latexsym}
\usepackage{graphics}
\usepackage{graphicx}
\begin{document}

\newcommand{\be}{\begin{equation}}
\newcommand{\ee}{\end{equation}}
\newcommand{\bea}{\begin{eqnarray}}
\newcommand{\eea}{\end{eqnarray}}
\newcommand{\wee}[2]{\mbox{$\frac{#1}{#2}$}}
\newcommand{\unit}[1]{\,\mbox{#1}}
\newcommand{\degree}{\mbox{$^{\circ}$}}
\newcommand{\ltish}{\raisebox{-0.4ex}{$\,\stackrel{<}{\scriptstyle\sim}$}}

\newcommand{\vs}{{\em vs\/}}
\newcommand{\bin}[2]{\left(\begin{array}{c} #1 \\
#2\end{array}\right)}
 
\title{Anisotropic velocity distributions in 3D dissipative optical
lattices}

\author{J.Jersblad\inst{1} \and H.Ellmann\inst{1} \and
L.Sanchez-Palencia\inst{2} \and A.Kastberg\inst{1,3}}
\institute{Department of Physics, Stockholm University, S-106 91
 Stockholm, Sweden \and
 Laboratoire Kastler-Brossel, D{\'e}partement
 de Physique de l'Ecole Normale Sup{\'e}rieure, 24 rue Lhomond, F-75231
 Paris cedex 05, France \and
 Department of Physics, Ume{\aa} University,
 S-901 87, Ume{\aa}, Sweden}
\date{Received: date / Revised version: date}

\abstract{We present a direct measurement of velocity distributions in two
dimensions by using an absorption imaging technique in a 3D near
resonant optical lattice. The results show a clear difference in
the velocity distributions for the different directions. The experimental
results are compared with a numerical 3D semi-classical Monte-Carlo
simulation. The numerical simulations are in good qualitative agreement with
the experimental results.}


\maketitle

\section{Introduction}
\label{intro}
An optical lattice is a periodic optical light shift potential created by the
interference of laser beams in which atoms can be trapped. 
Usually one distinguish between two types of lattices,
near-resonance optical lattices (NROL) \cite{mennerat02}
and far-off resonance lattices (FOROL). In the later type,
an atom can only be trapped, whereas the former
(the one considered in this paper)
also exhibits an inherent cooling mechanism (Sisyphus cooling).
The Sisyphus cooling mechanism in an NROL has been the
subject of extensive research due to its high cooling efficiency,
but also since an optical lattice is a very pure quantum system
suitable for fundamental studies of atom-light interaction.

Theoretical studies of the atomic motion in NROLs have been done in 1D
and 2D, both analytically and numerically.  The extension to 3D
configurations is however cumbersome. Analytical solutions become
unwieldy and numerical simulations require long computation time,
especially for high angular momentum transitions. Thus very few
detailed studies have been made in 3D. An exception is the work
by Castin and M\o lmer \cite{molmer95} who studied spatial and
momentum localization via full quantum Monte Carlo wavefunction
simulations in the case of optical molasses.

Measurements of temperature have been made on 3D NROLs by our group
\cite{ellmann,jersblad}, and by groups at NIST \cite{gatzke} and in
Paris \cite{mennerat}.  In all these experiments,
and in this work, the kinetic temperature is derived from measured velocity
distributions along one axis and is defined as a direct measure of the kinetic
energy through
\be
T=\frac{M\langle v^2\rangle}{k_{\textrm{\scriptsize B}}},
\label{temperature_def}
\ee
where $\langle v^2\rangle$ is the mean square velocity of the released
atoms, $k_{\textrm{\scriptsize B}}$
is the Boltzmann constant and $M$ is the atomic mass.
A robust result in all studies is that the temperature scales linearly
with the irradiance divided by the detuning, that is linearly with the
light shift at the bottom of the optical potential ($U$).  This is in
excellent qualitative agreement with 1D-theoretical predictions
\cite{dalibard}.

Nevertheless, in several works (for example \cite{ellmann},
\cite{gatzke} and this work) a four laser beam configuration results
in a face centered tetragonal lattice that cannot simply be reduced to
three 1D cases.  Indeed, all spatial directions are not equivalent
(see section~\ref{experimental}) and the particular geometry of the
lattice has to be taken into account.  In a recent paper by the
Grynberg group \cite{carminati}, the dependence of temperature and
spatial diffusion on geometric parameters controlling the lattice
spatial periods (lattice constants) in different directions was studied.
For different
laser beam configurations producing the NROL, the temperature and
spatial diffusion coefficient were measured for tetragonal lattices
 (see section~\ref{experimental})
with different aspect ratios, i.e. as a function of lattice constants. 
It was shown that the spatial diffusion coefficient strongly depends
on the direction. The temperature, which was measured in one
direction, was found to be independent of the lattice spacing.  The
difference between spatial directions lies not only in the lattice
constants, but also in the modulations of the laser-atom interaction
parameters (optical potentials and optical pumping) in such a way that
different behaviors of the temperature along different axes is
possible.  In \cite{palencia} the Sisyphus cooling effect in a 3D
tetragonal NROL was studied theoretically.  With a simplified choice
of atomic angular momentum, it was shown by a semi-classical
Monte-Carlo calculation that the temperature along a given coordinate
axis is independent of the lattice constant, but indeed different
along different directions.  For the same geometry as considered here,
the linear scaling parameter of the temperature differs by a factor of
1.4.  Moreover, a comparison between \cite{ellmann} and \cite{gatzke}
suggests such an anisotropy of the velocity distribution.  In both
experiments, the direction of measurement coincided with the direction
of gravity, but this direction did not correspond to the same lattice
axis.  It turns out that these works yield a quantitative discrepancy. 
The derived temperature was found to be linear with $U$ with
proportionality constants of 12 nK/$E_{\textrm{{\scriptsize R}}}$ and
24 nK/$E_{\textrm{\scriptsize R}}$ (in \cite{ellmann} and
\cite{gatzke} respectively), where $E_{\textrm{\scriptsize R}}$ is the
recoil energy \footnote{The recoil energy $E_\textrm{R}=(\hbar
k)^2/2M$, where $k=2\pi/\lambda$ is the wave vector, $\lambda$ is the
wavelength of the light, and $M$ is the atomic mass.}.  The
difference in scaling factor called out for a more thorough
investigation, which would rule out any systematic error.

This work aims at a direct comparison between the kinetic temperatures
along different directions in a 3D NROL. Measurements of velocity
distributions along different directions were made for different
lattice parameters (potential depth and detuning) by absorption
imaging of an expanding atomic cloud.  The experimental results are
compared with a 3D semi-classical Monte-Carlo simulation performed for
the actual atomic angular momentum.  

The paper is organized as follows.  In section~\ref{experimental} we
describe the experimental set-up.  The experimental data is presented
with derived kinetic temperatures in section~\ref{measure}.  In
section~\ref{theory}, we describe the numerical calculation and
present the result for the kinetic temperatures.  In section~\ref{discussion}
we discuss the results from the experiment and the simulations.
Finally, in section~\ref{conclusion} we draw conclusions on our work.

\section{Experiment}
\subsection{Experimental setup}
\label{experimental}
Initially, a magneto-optical trap (MOT) is loaded with ${N}\approx
2\cdot 10^6$ cesium atoms (\textsuperscript{133}Cs) from a chirped
decelerated atomic beam in 4 s.  This gives a peak number density of
$n_{0}\approx 5\cdot 10^{10}$ cm$^{-3}$.  The MOT operates at the
$({F}_{\textrm{g}}=4\rightarrow{F}_\textrm{e}=5)$ transition at 852 nm
(the D2 line), where $F$ is the total angular momentum quantum number.  Due to
off-resonant excitation to $F_\textrm{e}=4$, a repumper beam resonant with the
\mbox{($F_\textrm{g}=3\rightarrow{F}_\textrm{e}=4$)} transition is
also used.  After turning off the loading, the atoms are further
cooled in an optical molasses for about 20 ms.  From the optical
molasses, an atomic cloud at a temperature of $T=3$ $\mu$K is loaded
into the optical lattice with a transfer efficiency of about 50\%. 
The filling factor of the lattice is around 0.2 \%.  The optical
lattice beams are red detuned from the
\mbox{$({F}_\textrm{g}=4\rightarrow{F}_\textrm{e}=5)$} resonance,
typically between \mbox{$\Delta_{5}=-10\Gamma$} and
\mbox{$\Delta_{5}=-40\Gamma$}, where \mbox{$\Gamma/2\pi=5.2$ MHz} is
the natural linewidth.  The atoms equilibrate in the lattice for 10 ms
and are then released by turning off the optical lattice beams,
with an acousto-optical modulator (AOM) in  less than $1 {\mu}s$,
followed by a measurement of the kinetic temperature.
This short falltime of the AOM avoids adiabatic release of the atoms
in the optical lattice.

\begin{figure}
\begin{center}
\includegraphics[scale=0.35]{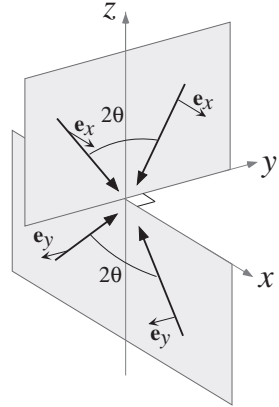}
\caption{Beam configuration of the 3D lin $\perp$ lin optical lattice.  Two
beam pairs in the $xz$- and $yz$-planes respectively, orthogonally
polarized along the $y$- and $x$-axes respectively, make an angle
$\theta=45^{\circ}$ with the $z$-axis.\label{stralkonfig}}
\end{center}
\end{figure}

\begin{figure}
  \begin{center}
  \includegraphics[scale=0.4]{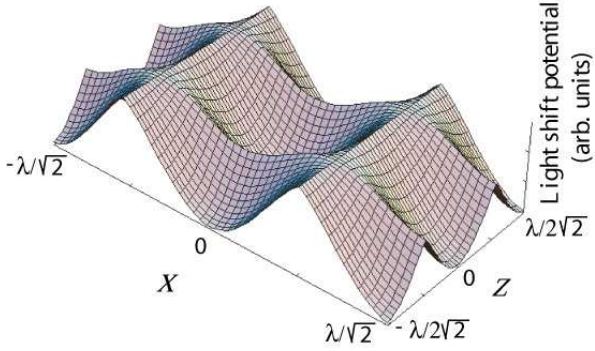}
  \caption{Lowest adiabatic 
  optical lightshift
  potential projected in the $xz$-plane in units of the optical
  wavelength, $\lambda$.\label{potential}}
  \end{center}
  \end{figure}

The optical lattice is a 3D generalization of the 1D lin$\perp$lin
configuration created by two orthogonally polarized pairs of laser
beams that propagate in the \textit{yz}- and \textit{xz}-planes
respectively \cite{mennerat02}. The angle between the beams of each pair is
$90^{\circ}$, and each beam forms an angle of
\mbox{$\theta=45^{\circ}$} with the (vertical) quantization
(\textit{z}-) axis (see figure \ref{stralkonfig}).  This results in a
tetragonal structure with alternating sites of pure $\sigma^{+}$- and
$\sigma^{-}$-light, where potential minima are formed.  From
figure~\ref{stralkonfig}, it is clear that directions \textit{x} and
\textit{y} are equivalent but that direction \textit{z} is different.
It follows that the optical pumping rates and the light shift
modulations are different along \textit{z} compared to \textit{x} or
\textit{y}.  In figure~\ref{potential} we plot the projection along
\textit{x} and \textit{z} of the lowest adiabatic potential¬, which is where the
atoms spend most of their time \cite{mennerat02}. Two main anisotropic
properties arise.  First, the lattice constants
$a_{z}=\lambda/(2\sqrt{2})$ and $a_{x,y}=\lambda/\sqrt{2}$ are
different.  Second, the shapes of the potentials are
also clearly different. In particular, they show
different potential barriers
to escape adiabatically from a potential well (lower along the
\textit{z}-direction than what it is along the \textit{x-} and
\textit{y-}directions by a factor of 1.65) and show different reduced oscillating
frequencies \mbox{($\omega_i a_i/\lambda$)} at the bottom of
the potential wells.

\begin{figure*}
\begin{center}
\includegraphics[scale=0.35]{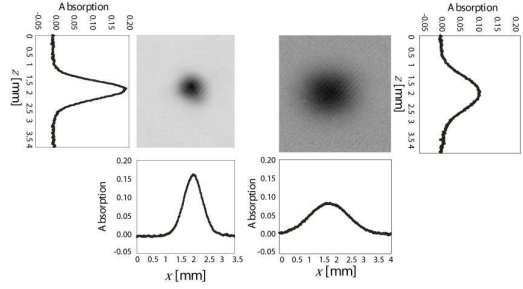}
\caption{\label{bilder} Typical 2D density profiles acquired at two
different times $\tau$ after releasing the atoms from the lattice.
The left image shows an atomic cloud after $\tau=12.8$ ms expansion
together with density profiles in the $z$- and $x$-directions.
The right image shows an atomic cloud after $\tau=36.8$ ms. }
\end{center}
\end{figure*}

The velocity distributions along the \textit{z}- and \textit{x}-axes
are measured using a well known absorption imaging technique
\cite{walhout}.  After release from the lattice, a short (50 $\mu$s)
resonant probe pulse
\mbox{$(F_\textrm{g}=4\rightarrow{F}_\textrm{e}=5)$} hits the atomic cloud.
The irradiance of the probe pulse is
$I \ll I_{0}$, where $I_{0}=1.1$ mW/cm$^{2}$ is the saturation
irradiance.  The shadow in the probe beam is
imaged onto a CCD camera.  By capturing images at different time
delays after turning off the optical lattice beams, we extract the
different spatial density distributions from which velocity
distributions can be derived.  The velocity distribution in the
\textit{z}-direction (direction of gravity) was compared to the results obtained with a
"time-of-flight" (TOF) method \cite{lett}, showing good agreement.

\subsection{Measured Kinetic Temperatures}
\label{measure}

The 2D projection (in the \textit{xz}-plane) of the expanding cloud is
recorded at two different time delays, $\tau_{1,2}$, after extinction
of the optical lattice beams.  Typical values are $\tau_{1}$ = 12 ms
and $\tau_{2}$ = 35 ms.  Examples of 2D density profiles are shown in
figure \ref{bilder} together with Gaussian fits to the spatial density
profile along \textit{x} and \textit{z}.  Excellent agreement with
Gaussian distributions is found.  From the fits, we extract the rms
radius, $\sigma_{i}$, ($i=x,z$), of the clouds which increases with
time, $t$, according to
$\sigma_i^{2}(t)=\sigma_{i}^{2}(0)+v_{i}^{2}t^{2}$ \cite{weiss}.

\begin{figure}
\begin{center}
\includegraphics[scale=0.4]{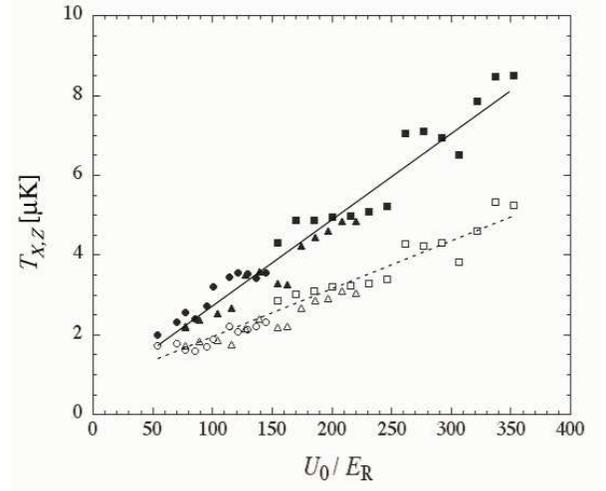}
\caption{\label{expresult} $T_{x}$ (filled) and
$T_{z}$ (open) as a function of modulation depth, $U_{0}/E_{R}$,
for three different detunings ($\Delta_{5}=-10\Gamma$ (squares)
, $-20\Gamma$ (triangles), $-30\Gamma$ (circles)).
The solid and dashed line are linear fits to the data.}
\end{center}
\end{figure}

The kinetic temperature in different directions is defined as
\be
T_{i}=\frac{M}{k_{\textrm{\scriptsize B}}}
\frac{\sigma_{i}^{2}(\tau_2)-\sigma_{i}^{2}(\tau_1)}{\tau_{2}^{2}-\tau_{1}^{2}}.
\label{temp}
\ee
In figure \ref{expresult} we plot derived kinetic temperatures
along \textit{x} and \textit{z} for three different detunings, as a
function of $U_{0}$, which is the modulation depth of the diabatic
optical potential.  Here, $U_{0}$ is defined as
\be
U_{0}=\frac{\hbar|\Delta_{5}|}{2}\ln\left[1+
{\left(\frac{44}{45}\right)\frac{\Omega^{2}}{2\Delta_{5}^{2}}}\right],
\ee\label{dibat}
where $\Omega^{2}=(\Gamma^{2}/2)/(I/I_{0})$ is the
square of Rabi frequency and the irradiance is
$I=8I_{\textrm{\scriptsize beam}}$ ($I_{\textrm{\scriptsize beam}}$ is
the irradiance of a single beam), at the center of a potential well.
For sufficiently high irradiances and temperatures, it is obvious that
the universal scaling with $U_{0}$ prevails for each direction.  
However, this scaling with $U_{0}$ is
clearly different for different directions.  
For large $U_{0}$, the temperature along \textit{z} is found to be
significantly smaller than the temperature along \textit{x}.
Linear fits to the data yield
\bea
T_{x}=\left(0.55+0.022
({U_{0}}/{E_{\textrm{R}}})\right) &\mu \textrm{K} \\
T_{z}=\left(0.62+0.012 ({U_{0}}/{E_{\textrm{R}}})\right) &\mu
\textrm{K}.
\eea
That is, the ratio between the scaling parameters
along \textit{x} and \textit{z} is determined to be 1.8 (0.3). However,
at low modulation depths and low temperatures, the
temperatures are found to be approximately the same along \textit{z} 
and \textit{x}.

\section{Numerical Simulations}
\label{theory}
\subsection{Theoretical framework}
We have performed
semi-classical Monte-Carlo simulations in 3D for the actual
\mbox{$(F_\textrm{g}=4\rightarrow{F}_\textrm{e}=5)$} transition of $^{133}$Cs.
The main features of the
method have been discussed elsewhere \cite{palencia,petsas99}
so here we just recall the main elements and peculiarities for
our multidimensional configuration.

The optical Bloch equations (OBE), which describe
the evolution of a sample of two-level atoms (with Zeeman degeneracy)
coupled to both laser fields and vacuum modes, are the starting point of the
analysis. Because
of the cooling effects and
the decoherence due to photon scattering, the atomic cloud
dynamics can be reduced to a semi-classical picture for a large range
of lattice parameters \cite{cohen90}. The OBE are therefore converted into
a set of coupled semi-classical Fokker-Planck equations (FPE)
via Wigner transforms. Projecting the FPE onto the
position-dependent adiabatic states base $|\Phi_m
(\mathbf{r})\rangle$ (see appendix \ref{bloch}) and
neglecting the coherence terms which are unimportant in a semi-classical
description, one gets a new set of FPE only involving the local
populations of the adiabatic states\footnote{Note that
the adiabatic approximation is justified
by the fact that the adiabatic state splittings are generally greater
than the motional couplings in the regime of deep potentials.}.

By physical interpretation of the FPE, it follows that the atomic cloud
dynamics can be reduced to internal state transitions via optical pumping at a rate
$\gamma_{n,m}$ from $|\Phi_n\rangle$ to $|\Phi_m\rangle$,
and the evolution of each atom in a given internal $|\Phi_m\rangle$-state due
to deterministic forces.
These forces are first of all due to the optical potential modulation
(-\mbox{\boldmath $\nabla$}$ U_m$) and secondly, due to the
radiation pressure force ($\mathbf{F}$).
Moreover, the atomic cloud undergoes momentum diffusion due to photon
scattering.

It is then straightforward to show that the FPE solution is formally
equivalent to the integration of a set of Langevin equations
interrupted by internal states quantum jumps, each one accounting for
the random trajectory of a single atom.
The quantum jumps are taken into account by generating a random
number $r$ at each time step which is compared
to the transition probability $\gamma_{m,n}\textrm{d}t$ from $|\Phi_m\rangle$
to $|\Phi_n\rangle$ (with $n \neq m$) during the time step d$t$.
In the following, we define $r_{n,m}$ as $1$ if a quantum jump occurs
from $n$ to $m$ and $0$ otherwise.
Between two quantum jumps, the elementary evolution of the atom
is
\bea
 \textrm{d}\mathbf{R}\left(t\right) & = &
\frac{\mathbf{P}\left(t\right)}{M} \textrm{d}t \\
 \textrm{d}\mathbf{P}\left(t\right) & = &
-\mbox{\boldmath $\nabla$} U_m \textrm{d}t
 + \sum_{n \neq m} r_{n,m} \left( \mathbf{\delta p}_{n,m}
 + \mathbf{F}_{n,m} \textrm{d}t \right) \nonumber \\
 & & + \big( 1-\sum_{n \neq m} r_{n,m} \big) \left(
\mathbf{f}_{m} + \mathbf{F}_{m,m} \right) \textrm{d}t,
\eea \label{langevin}
where $\mathbf{R}$ and
$\mathbf{P}$ are the atomic position and
momentum respectively.
The Hamiltonian force, (-\mbox{\boldmath $\nabla$}$U_m$), is 
derived from the adiabatic potential in state $|\Phi_m\rangle$, and
$\mathbf{F}_{n,m}$ is the average radiation pressure
in case of a quantum jump from $n$ to $m$ (if $m=n$, no jump occurs).
The momentum diffusion is determined by random values: the momentum kick
undergone by the atom in case of a quantum jump from $n$ to $m$,
$\mathbf{\delta p}_{n,m}$ and the recoil mean force in the
absence of a quantum jump, $\mathbf{f}_m$. Note that
$\mathbf{\delta p}_{n,m}$
and $\mathbf{f}_m$ are related to the position-dependent
coefficients appearing within the FPE. In a (${\mu}$-indexed) space base where the
momentum diffusion
matrix $\{D_{n,m}\}$ (see appendix \ref{coefficients})
is diagonal, the first two moments of
$\mathbf{f}_m$
and $\mathbf{\delta p}_{n,m}$ read
\bea
 \langle f_m^\mu \rangle = 0 & \textrm{   and   } & \langle ( f_m^\mu
) ^2 \rangle = \frac{2 D_{m,m}^{\mu,\mu} \left( \mathbf{r}
\right)}{dt} \nonumber \\
 \langle \delta p_{n,m}^\mu \rangle = 0 & \textrm{   and   } &
\langle (\delta p_{n,m}^\mu ) ^2 \rangle = \frac{2 D_{n,m}^{\mu,\mu}
\left( \mathbf{r} \right)}
 {\gamma_{n,m}}.
\eea \label{kicks}

\subsection{Numerical results}
\begin{figure}
\begin{center}
\includegraphics[scale=0.40]{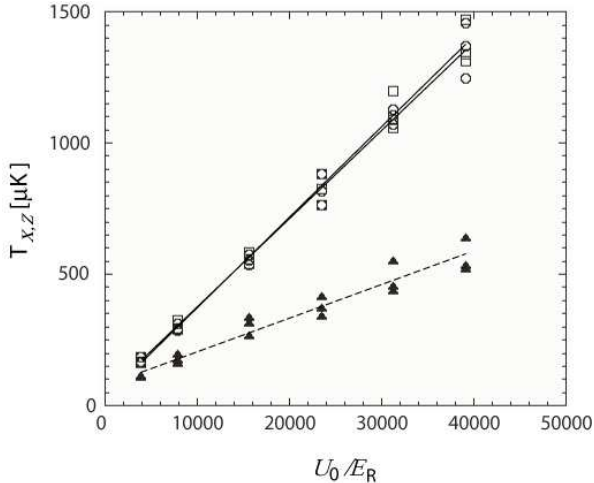}
\caption{\label{numresult} Kinetic temperature, along the $x$- (circles), $y$-
(squares) and
$z$-direction (filled triangles) as a function of modulation depth,
$U_\textrm{\scriptsize 0}/E_\textrm{\scriptsize R}$,
for three different detunings ($\Delta_{5}=-10\Gamma,-20\Gamma,-30\Gamma$).
The solid and dashed lines are linear fits to the data.}
\end{center}
\end{figure}

The numerical simulations are performed for a typical sample of $300$
independent atoms.  For the lattice parameters considered in this
work, the kinetic energy reaches steady-state in a time of approximately
$4000/\Gamma^{'}$, where $\Gamma^{'}=\Gamma s_{0}/2$ is the total scattering rate
and $s_{0}$ is the saturation parameter (see appendix \ref{bloch}). 
The averages of the kinetic energies in steady
state in the $x$-, $y$- and $z$-directions provide the kinetic temperatures
in the corresponding directions,
\be
T_{i}=\frac{M\langle v_i^{2} \rangle}{k_{\textrm{\scriptsize B}}}.
\ee
\label{temp_simul}
The simulations were made for three different detunings
($\Delta_{5}=-10\Gamma, -20\Gamma, -30\Gamma$).  For each detuning we
acquired velocity distributions, in each direction, at six different
modulation depths.  Note that the chosen modulation depths are much
higher than in the experiment since the semi-classical model breaks
down when the momentum distribution becomes too narrow. This is because deep
modulation depths are required to avoid non-adia\-ba\-tic motional
couplings between adiabatic sublevels that are not included in our treatment
\cite{petsas99}.  Moreover, the
time to reach steady state increases for low modulation depths. 
However, the linear scaling should still hold.  The results of the
numerical simulations are shown in figure \ref{numresult}.  Here, the
kinetic temperature is plotted as a
function of modulation depth for the detunings mentioned above.  The
temperature scales linearly with the light shift independently of
the detuning according to
\bea
   T_{x}\propto 0.035 (U_\textrm{\scriptsize 0}/E_\textrm{\scriptsize R}) &\mu \textrm{K}
\\ T_{y}\propto 0.035 (U_\textrm{\scriptsize 0}/E_\textrm{\scriptsize R}) &\mu \textrm{K}
\\ T_{z}\propto 0.013 (U_\textrm{\scriptsize 0}/E_\textrm{\scriptsize R}) &\mu \textrm{K}
\eea
As in the experiments, the results of the simulations 
show a clear difference
in scaling of the kinetic temperature along the \textit{z}-axis compared to the \textit{x}-
and \textit{y}-axes, here, by a factor of 2.7.

\section{Discussion}
\label{discussion}
The results from the experimental work and the numerical simulations
are compiled in table \ref{tab:1}.  A comparison shows 
a quantitative excellent 
agreement between our experiments and former studies
in which the kinetic temperature was measured along \textit{x} \cite{gatzke}
or along \textit{z} \cite{ellmann}. The numerical simulations also
reproduce the difference in scaling parameter for different
directions was measured in the experiments, and
confirms the appearance of a discrepancy
between kinetic temperatures along \textit{x}-\textit{y} and \textit{z}.

\begin{table}[tbp]
    \begin{footnotesize}
\caption{The scaling parameter
\mbox{$\xi_{x,y,z}$}, (in units of nK/$E_{\textrm{R}}$), in the equation
\mbox{$T_{x,y,z}=T_{0}+\xi_{x,y,z}U_{0}$} for different studies.
The experimental
errors of the slope for this work is the quadratic sum of the statistical error
and an estimated maximum systematical error. The errors in the simulation is 
the statistical error from the fit.}
\label{tab:1}
\begin{tabular}{lllll}
\hline\noalign{\smallskip}
 & ref \cite{ellmann}
 & ref \cite{gatzke}
 & this work
 & this work  \\
 &
 &
 & (experimental)
 & (simulations)  \\
\noalign{\smallskip}\hline\noalign{\smallskip}
 $\xi_{x}$& - & 24(2.4) 
 & 22(3.5) 
 &
        35(1.2)  
	\\
        $\xi_{y}$& - & - & - & 35(1.2) 
	\\
	$\xi_{z}$& 12(1.2) 
	& - & 12(2.5)  
	& 13(1.0) 
	\\
\noalign{\smallskip}\hline
\end{tabular}
\end{footnotesize}
\end{table}

The inherent cooling process in an optical lattice for atoms  with 
kinetic energy $E_{\textrm{{\scriptsize K}}}>U_{0}$ 
is Sisyphus cooling. This process was explained by  Dalibard and
Cohen-Tannoudji in \cite{dalibard} in the case of a theoretical
transition $(J_{g}=1/2 \rightarrow J_{e}=3/2$). 
The Sisyphus cooling cycle occurs until the atomic kinetic energy 
is lower than the potential barrier in a particular direction and thus does not 
depend on any other anisotropy (the lattice spacings for example).

However, for higher angular momentum transitions, the cooling process
does not stop because other relaxation processes than
standard Sisyphus cooling could still occur \mbox{\cite{petsas99,deutsch}}.
For example, atoms in bound states within a lattice well
can be excited to unbound states,
followed by  decay to lower lying vibrational states.
We find that the atomic
kinetic energy is \mbox{$E_\textrm{\scriptsize K} \sim U_0/10$}
and thus that the atoms are
very well localized at the bottom of the lattice wells in agreement
with former experimental investigations, for instance \cite{gatzke},
and full quantum Monte-Carlo simulations \cite{molmer95}.
The difference in the
scaling factors is proportional to the
difference in the modulation depth of the lowest adiabatic optical 
potential in the corresponding directions. Therefore we conclude
that it is this difference which induce anisotropic kinetic temperatures in the optical
lattice.  This conclusion is not incompatible with the results of
\cite{carminati} in which the steady-state kinetic temperature was measured
for different lattice constants showing that the steady-state
kinetic temperature was independent of the lattice spacing, because the
geometrical anisotropy in the lattice do not reduce to a simple
scaling factor between directions \textit{x,y} and \textit{z}.

At low modulation depths, the lattice reaches a minimum temperature
followed by a sharp increase in temperature, usually called
d\'ecrochage.  When laser cooling is still effective there exists a
region where the temperature is isotropic.
However, this region is difficult to analyze for several reasons.
For instance, at low modulation depths the atomic
localization in a trapping site is less strong, and thus the
anharmonicity of the potential well becomes more important. This could lead to
an increased coupling between the different motional directions and
also a broadening of the vibrational levels, i.e. increasing the
tunneling rate in the lattice. Another effect that must be taken into
account at low modulation depths is increased spatial diffusion
\mbox{\cite{carminati,hodapp}}.
This means that the loss rate of the atoms in the lattice
becomes larger, and thus the signal-to-noise in the absorption images
decreases.
Furthermore, if the thermal expansion of the atomic cloud in the
recorded absorption images is small compared to the size of the cloud,
due to spatial diffusion, there will be large uncertainties in the
extracted temperatures.

\section{Conclusions}
\label{conclusion}
We have measured the velocity distributions in a 3D optical lattice of
cesium along two non-equivalent directions as a function
of lightshift ($U_0$). In agreement
with previous works, the kinetic temperature scales linearly with $U_0$.
As an original result, we have found that
the distributions are clearly anisotropic (with $T_{x,y}>T_z$).  The
experimental results are in good agreement with a 3D numerical
Monte-Carlo simulation and we conclude that it is the modulation depth
of the adiabatic optical potential that determines the steady-state
kinetic temperatures.
The anisotropy in kinetic temperature is not paradoxical. In fact
the "kinetic temperature" here is defined as a simple measure of the atomic kinetic energy
(see Eq.~(\ref{temperature_def})) and not as a thermodynamical temperature.
This is because thermalization in Sisyphus cooling do not result from energy exchange
beetween particles via collisions, but from atom-photon interactions.
Our result show that no thermodynamical temperature can be defined
for Sisyphus cooled atomic samples because of the violation of the equipartition theorem
\cite{landau80}. Our results can give important clues for a full understanding of the
cooling mechanism in an optical lattice. Furthermore, knowledge about the
velocity distributions in all directions is important in precision
experiments utilizing optical lattices.

\begin{acknowledgement}
LSP thanks the swedish group for warm hospitality during the period
when a part of this work was achieved.  He also acknowledges financial
support from the Swedish Foundation for International Cooperation in
Research and Higher Education (STINT).  We would like to thank Dr. 
Peter Olsson at Ume{\aa} University for letting us use the LINUX
cluster and also for support during the simulations at the theoretical
physics department at Ume{\aa} University.  This work was supported by
the Swedish Natural Sciences Research Council (NFR), the Carl Trygger
Foundation, the Magnus Bergwall Foundation and the Knut \& Alice
Wallenberg Foundation.
\end{acknowledgement}

\begin{appendix}
\section{Optical Bloch equations and adiabatic states}
\label{bloch}

This appendix aims at introducing the adiabatic states for a general
$J \rightarrow J+1$ transition atomic sample. Consider an
atom of dipole operator
$\widehat{\mathbf{D}} = \mathcal{D} (
\widehat{\mathbf{d}}^+ + \widehat{\mathbf{d}}^- )$
, with $\widehat{\mathbf{d}}^\pm$ being
the raising and lowering components of
$\widehat{\mathbf{D}}$, and $\mathcal{D}$ the reduced
dipole moment. This atom interacts with the laser field
\be
\mathbf{E}_\textrm{\scriptsize L} \left( \mathbf{r} , t \right) =
\frac{E_0}{2} \mathbf{\epsilon} \left( \mathbf{r} \right)
e^{-i\omega_{\textrm{\tiny L}} t} + c.c.
\ee
\label{field}
, where $E_{0}$ is the amplitude of the electric
field, $\omega_\textrm{\scriptsize L}$ is the laser frequency and
$\mathbf{\epsilon}(\mathbf r)$
is a vector describing the spatial varying profile of the laser polarization.
The operators $\widehat{A}$ and $\widehat{B}_q$
represent the hermitian conjugates of the optical pumping cycles (absorption
of laser photons followed by emission of stimulated or spontaneous photons
respectively),
and are defined as
\bea
& & \widehat{A}\phantom{_q} = \left[ \widehat{\mathbf{d}}^-
\cdot \mathbf{\epsilon}^* (\mathbf{r}) \right] \cdot
\left[ \widehat{\mathbf{d}}^+
      \cdot \mathbf{\epsilon} (\mathbf{r}) \right]
\nonumber \\
& & \widehat{B}_q = \left[ \widehat{\mathbf{d}}^- \cdot
\mathbf{\epsilon}^* (\mathbf{r}) \right] \cdot \left[
\widehat{\mathbf{d}}^+
      \cdot \mathbf{e}_q \right] \\
& & \textrm{with } q = 0, \pm \phantom{a} \textrm{or } q = x,y,z
\eea
\label{ab}
where
\be
\mathbf{e}_\pm = \frac{\mp \mathbf{e}_x - i\mathbf{e}_y}
        {\sqrt{2}} \phantom{aaa} \textrm{and}
        \phantom{aaa} \mathbf{e}_0 = \mathbf{e}_z
\ee \label{eq}
are the circular basis vectors. After elimination of the excited state
in the low saturation regime,
\be
s_0 = \frac{\Omega^2/2}{\Delta^2+\Gamma^2/4} \ll 1,
\ee\label{saturation}
the atomic sample dynamics is
governed by the OBE involving the projection of the density matrix onto
the internal state including an Hamiltonian
part:
\be
\widehat{H} = \frac{\widehat{\mathbf{p}}^2}{2 M} + \hbar
\frac{\Delta s_0}{2} \widehat{A} \left( \mathbf{r} \right)
\ee \label{hamiltonian}
plus a relaxation part. In the semi-classical limit, the
position-dependent adiabatic states are defined as the eigen-states
of the light-shift operator
$\hbar \frac{\Delta s_0}{2} \widehat{A}$:
\be
\hbar \frac{\Delta s_0}{2} \widehat{A} \left( \mathbf{r}
\right) | \Phi_m \left( \mathbf{r} \right) \rangle =
        U_m \left( \mathbf{r} \right) | \Phi_m \left(
\mathbf{r} \right) \rangle.
\ee \label{adiabatic}
Note that in general $| \Phi_m \left( \mathbf{r} \right) 
\rangle$ and $U_m \left( \mathbf{r} \right)$ cannot be 
calculated analytically.

\section{Dynamics coefficients for the Langevin equation}
\label{coefficients}
In this appendix, we give the general expressions for the dynamics 
coefficients involved in the FPE and Langevin equations for Sisyphus 
cooling in the low saturation
and semi-classical regime. The transition rate from state
$|\Phi_n\rangle$ to state $|\Phi_m\rangle$ (for $m \neq n$) is
\be
 \gamma_{n,m} = \Gamma_0' \sum_{q=\pm,0} | \langle \Phi_n | 
\widehat{B}_q | \Phi_m \rangle |^2.
\ee \label{feeding}
The average radiation pressure term in the direction \textit{i} (\textit{i}$=x,y,z$) is
\be
F_{n,m}^{i} = -\hbar \Gamma_0' \;\mathrm{Im}\left( \sum_{q=\pm,0} \langle 
\Phi_m | \partial_{i} \widehat{B}_q^\dagger | \Phi_n \rangle
                   \langle \Phi_n | \widehat{B}_q | \Phi_m \rangle 
\right)
\ee \label{radiation}
and the momentum diffusion matrix is
\bea
D_{n,m}^{i,j} & = & \frac{\hbar^2 \Gamma_0'}{8} \langle \Phi_n | 
\partial_{i,j}^2 \widehat{A} | \Phi_m \rangle \delta_{n,m} 
\nonumber \\
& & + \frac{\hbar^2 k^2 \Gamma_0'}{4} \delta_{i, j}
\sum_{\begin{array}{c} \phantom{a}^{u \in {x,y,z}} \\ \phantom{a}^{u \neq i,j}
\end{array}} \langle \Phi_m | \widehat{B}_u^\dagger | \Phi_n \rangle
       \langle \Phi_n | \widehat{B}_u | \Phi_m \rangle \nonumber \\
& & - \frac{\hbar^2 \Gamma_0'}{8} \sum_{q=\pm,0} \Big( \langle \Phi_m
| \partial_{i,j}^2 \widehat{B}_q^\dagger | \Phi_n \rangle
        \langle \Phi_n | \widehat{B}_q | \Phi_m \rangle \nonumber \\
& & \phantom{aaaaaaaaaaaa}- \langle \Phi_m | \partial_i 
\widehat{B}_q^\dagger | \Phi_n \rangle \langle \Phi_n | \partial_j
\widehat{B}_q | \Phi_m \rangle \nonumber \\
& & \phantom{aaaaaaaaaaaa}+ c.c. \Big)
\eea
\label{diffusion}
where $\delta_{\alpha, \beta}$ is the Kronecker symbol ($1$ when
$\alpha = \beta$ and $0$ else) and $i,j$ denotes the spatial directions
$(x,y,z)$.
Note that for the sake of simplicity, 
the spontaneous emission
pattern is simplified in a way that the photons are restricted to be 
emitted only along the $x$-, $y$- and $z$-axes. This approximation is 
justified because the kinetic energy is expected to be greater than the recoil energy 
\cite{castin94}.
\end{appendix}

\end{document}